\documentclass[twocolumn, times, tighten, appendixfloats]{aastex631}

\usepackage{natbib}
\usepackage{enumitem}
\usepackage{amsmath}
\usepackage{soul}
\usepackage{longtable}

\usepackage{graphicx}		% for including figures
\usepackage{latexsym}		% additional math symbols
\usepackage{bm}			% bold math package
\usepackage[english]{babel}
\usepackage{color}		% for colored text and boxes; remove for subm.
\begin{document}

\title{Point-like sources among $z>11$ galaxy candidates: contaminants due to supernovae at high redshifts?}

\author[0000-0001-7592-7714]{Haojing Yan} % yanhaojing@gmail.com
\affiliation{Department of Physics and Astronomy, University of Missouri, Columbia, MO\,65211, USA}
%\email{yanha@missouri.edu}

\author[0000-0001-7092-9374]{Lifan Wang}%
% \email{lifan@tamu.edu}
\affiliation{George P. and Cynthia Woods Mitchell Institute for Fundamental Physics \& Astronomy, \\
Texas A. \& M. University, Department of Physics and Astronomy, 4242 TAMU, College Station, TX 77843, USA}

\author[0000-0003-3270-6844]{Zhiyuan Ma} %\email{zhiyuanma@umass.edu}
\affiliation{Department of Astronomy, University of Massachusetts, Amherst, MA 01003, USA}

\author[0000-0001-7201-1938]{Lei Hu}
%   \email{hulei@pmo.ac.cn}
\affiliation{Purple Mountain Observatory, Chinese Academy of Sciences, No. 10 Yuanhua Road, Qixia District, Nanjing, Jiangsu 210023, China}

\begin{abstract}

   The recent searches for $z>11$ galaxies using the James Webb Space Telescope
have resulted in an unexpectedly high number of candidate objects, which imply
at least an order of magnitude higher number density of $z>11$ galaxies than
the previously favored predictions. A question has risen whether there are some
new types of contaminants among these candidates. The candidate sample of 
Yan et al. (2023a), totalling 87 dropouts, is the largest one, and we notice that
a number of these candidates are point-like. We hypothesize that the 
point-source dropouts could be supernovae at high redshifts. Further 
investigation shows that most of their spectral energy distributions indeed can 
be explained by supernovae at various redshifts from $z\sim$ 1--15, which lends
support to this hypothesis. Attributing such point-source dropouts to
supernova contamination cannot eliminate the tension, however, because they only
account for $\sim$10\% of the Yan et al.'s sample. On the other hand, the
discovery of ``contaminant'' supernovae at $z>3$ will have a series of important 
implications. Ironically, the existence of supernovae at $z>10$ would still 
imply that the previously favored picture of early galaxy formation severely 
underestimates the global star formation rate density such redshifts. Multiple-
epoch JWST imaging  will be the simplest and yet the most efficient way to 
further test this hypothesis.

\end{abstract}

\section{Introduction}

   Since its Early Release Observations (ERO), James Webb Space Telescope (JWST) 
has created a flux of new searches for high-redshift ($z>11$) candidate galaxies
\citep[][]{Naidu2022a, Castellano2022, Atek2023, Finkelstein2022b,  
Harikane2022, Bouwens2022, Adams2023, Donnan2023, Rodighiero2023, Yan2023a, 
Yan2023b}. 
The large number of candidates (over a hundred as of this writing) resulted from 
these studies are incompatible with the previously favored predictions at
$z>11$ \citep[see e.g.,][]{Behroozi2020, Vogelsberger2020}. The problem is 
exacerbated by the fact that some of these candidates are 
much brighter ($m\lesssim 26.5$~mag) than what would be expected for galaxies at 
such high redshifts \citep[see e.g.,][for a summary]{Yan2023b}. If a significant 
fraction of these objects are indeed at $z>11$, we will have a severe difficulty 
to reconcile them to our current picture of galaxy formation in the early 
universe.

   Among all the $z>11$ candidate samples selected using the early JWST data,
the one presented by \citet[][hereafter Y23]{Yan2023a} is the largest and has
attempted to probe the highest redshift range. It was based on the 6-band NIRCam
images of the nearby galaxy cluster SMACS J0723-73, which were taken as part of 
the JWST ERO \citep[][]{Pontoppidan2022}. 
Only half of the field centered on the cluster is boosted by lensing, and the 
other half is not affected. Their candidates were selected as the dropouts from 
F150W (60 objects), F200W (15 objects) and F277W (12 objects), respectively, 
which nominally correspond to $z\approx 12.7$, 17.3, and 24.7, respectively. 
Such a large number of $z>11$ candidate objects pose the most severe challenge 
to the current picture of galaxy formation in the early universe.

   Y23 also cautioned that some of their candidates,
while having good SED fits consistent with being at $z>11$, could still be due
to some new types of contaminators that were not encountered previously in  
high-$z$ search. We notice that a few of these dropouts are point-like 
sources. In the dropout searches at lower redshifts ($z\leq 10$), such point-like
sources would be identified as contaminants due to Galactic brown dwarf stars. 
However, Y23 demonstrated that their colors are inconsistent with brown dwarfs,
which can be understood because the broad molecular absorption bands of
brown dwarfs do not locate in the wavelength range of our interest.
All this motivates us to consider if any of these point sources could be 
supernovae (SNe). SNe exhibit a wide range of properties in their SEDs at 
different evolutionary stages, and they generally have a sharp cut-off at 
rest-frame $\lesssim 4000$\AA\ post-maximum because of the sudden onset of 
strong metal line absorption (see \S~\ref{sec:SNInter} for the details of the
SED fits). This raises the possibility that they could become contaminators to
high-$z$ searches. In this Letter, we explore such a possibility using the 
slightly extended dropout sample of Y23. The selection of point sources in this 
sample is given in Section 2. We study the supernova (SN) interpretation
by carrying out SED fitting to the model SN templates, which is detailed in
Section 3. We discuss the implication of our results in Section 4 and conclude
with Section 5. All magnitudes quoted are in the AB system. All coordinates are 
of J2000.0 Equinox. We adopt the following cosmological parameters:
$\Omega_M=0.27$, $\Omega_\Lambda=0.73$ and $H_0=71$~km~s$^{-1}$~Mpc$^{-1}$.

\section{\label{ptsource}Point sources among dropouts} 

    The JWST ERO NIRCam observations of SMACS 0723-73 were done in six broad
bands, namely, F090W, F150W and F200W in the ``short wavelength'' (SW)
channel and F277W, F356W and F444W in the ``long wavelength'' (LW) channel.
Hereafter we refer to the magnitudes in these bands as $m_{090}$, $m_{150}$,
$m_{200}$, $m_{277}$, $m_{356}$, and $m_{444}$, respectively.
Y23 selected $z>11$ candidates using the standard
``dropout'' technique \citep[][]{Steidel1995} to identify the characteristic 
Lyman-break signature in their SEDs. They adopted the color decrement of
$\geq 0.8$~mag and signal-to-noise ratio S/N $\leq 2$ (i.e., non-detections)
in the veto band(s). For reference, when the break moves to halfway of the 
drop-out band, the color decrement is $\sim$0.75~mag if the SED is flat in
$f_\nu$. In total, they have selected 60 F150W dropouts, 15 F200W dropouts and 
12 F277W dropouts, respectively. This is the largest high-$z$ dropout sample 
to date.

   Our visual inspection of these dropouts suggests that
a number them are point-like. To confirm that they are indeed point sources, we
carried out further investigations. The NIRCam image stacks used in Y23
were all produced at the pixel scale of 0\arcsec.06 (hereafter
``60mas''), which was chosen to match the native pixel scale of F356W and to
optimize the detection of faint objects. However, such a scale is not
ideal for our purpose here because it undersamples the three SW bands (F090W, 
F150W and F200W) passbands, which have better resolutions (as much as 
$\sim$2$\times$) than the three LW bands(F277W, F356W and F444W). Therefore, we
created a new version of images at the pixel scale of 0\arcsec.03
(hereafter ``30mas'') to identify point sources. Thanks to the sufficient 
dithers employed by this set of observations, the 30mas stacks are critically 
(Nyquist) or better sampled in SW and oversampled in LW.
We also updated the reference files to the JWST calibration reference data system
(CRDS) context ``jwst\_1008.pmap'', which incorporates the best flux zero-point 
calibration as of this writing. To take the full advantage of the wavelength 
coverage of the NIRCam data, we extended the Y23 dropout sample to include 
F356W dropouts. These were selected from a source catalog constructed in a
similar way as in Y23, with the difference that the F444W images
were used for the source detection and aperture definition. A legitimate F356W
dropout must have S/N $\geq 5$ in F444W, satisfy $m_{356}-m_{444}\geq 0.8$~mag, 
and have S/N $\leq 2$ in all the veto bands (i.e., those bluer than F356W). To
ensure sufficient S/N in evaluating their sizes, we truncated the F356W dropouts
to $m_{444}\leq 29.0$~mag. As it turns out, there is only one such F356W dropout
left after this truncation.

   To identify the point sources among the dropout sample, we first singled
out those that appears to be point-like by visual inspection in F200W, F277W,
F356W, and F444W. We then fitted a 2-D gaussian profile to these objects to 
check whether they had the full-width-at-half-maximum (FWHM) values consistent
with the expectations for point sources seen by the NIRCam. As the in-flight 
NIRCam PSFs have not yet been fully characterized in this early stage of the JWST
mission, we took an empirical approach, which is detailed in Appendix. Briefly,
we selected high S/N point sources in the field based on the diagnostics used
by the PSFEx software tool \citep{Bertin2011}, determined the average FWHM 
values and the dispersions in the four aforementioned bands, and compared the 
FWHM values of our dropouts to these statistics. A dropout is deemed a point 
source if its FWHM in at least one band is within 5~$\sigma$ of the average in 
this band. As an extended source can never have FWHM of a point source, this one-
band requirement will not include non-point sources. We did not impose this 
requirement to all bands where the object is detected, as it would be too 
stringent. First of all, a point source could have a large FWHM measured when 
lacking sufficient S/N. Second, the drizzle algorithm \citep{Fruchter2002} used 
by the JWST image stacking routine does not strictly preserve the source FWHM 
\citep{Fruchter2011}. Therefore, if a source has a point-like FWHM in one band, 
it is safe to conclude that it is a point source.

\begin{figure*}[htbp]
    \centering
    \includegraphics[width=\textwidth]{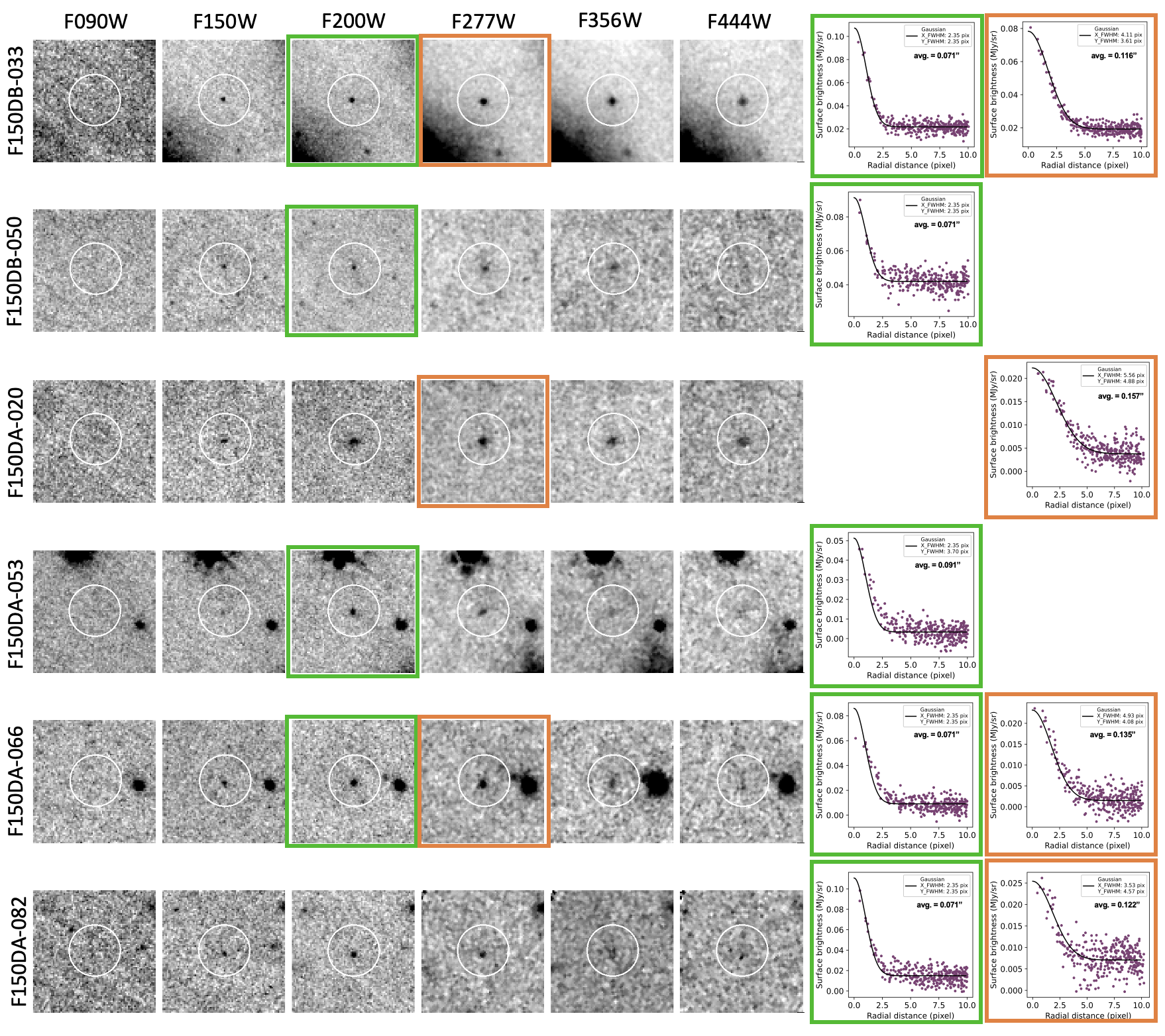}
    \caption{Six-band image stamps and profile fitting results of the six point-
    source F150W dropouts in the Y23 sample (among 60 in total). The image 
    stamps (2\arcsec.4$\times$ 2\arcsec.4 in size) are based on the 30mas 
    stacks, and the white circles (0\arcsec.5 in radius) are centered on the 
    dropouts. The dropout ``short IDs'' (as in Y23) are labeled to left, while 
    the passbands are labeled on top. The images used for the 2-D gaussian 
    profile fitting are outlined by the green (for F200W) and orange (for F277W) 
    boxes. The corresponding fitting results are shown in the boxes to right, 
    outlined by the same colors. The legends show the image FWHM values in units 
    of pixels along both axes, and the averages in units of arc seconds are also 
    labeled.   
}
    \label{fig:pts_f150d}
\end{figure*}

\begin{figure*}[htbp]
    \centering
    \includegraphics[width=\textwidth]{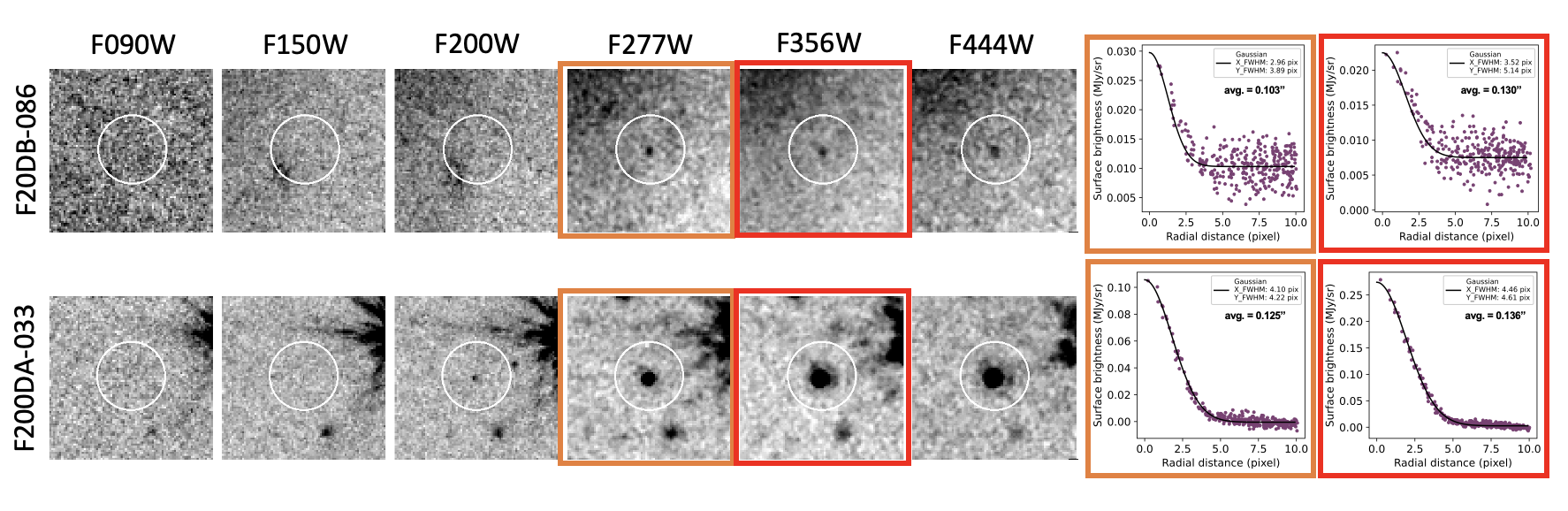}
    \caption{Similar to Figure \ref{fig:pts_f150d}, but for the two point-source
    F200W dropouts in the Y23 sample (among 15 in total). The images used for 
    the profile fitting are outlined by the orange (for F277W) and red (for 
    F356W) boxes. F200DA-033 also satisfies the point-source criteria in F444W, 
    which is not shown due to the limited space.
}
    \label{fig:pts_f200d}
\end{figure*}

\begin{figure*}[htbp]
    \centering
    \includegraphics[width=\textwidth]{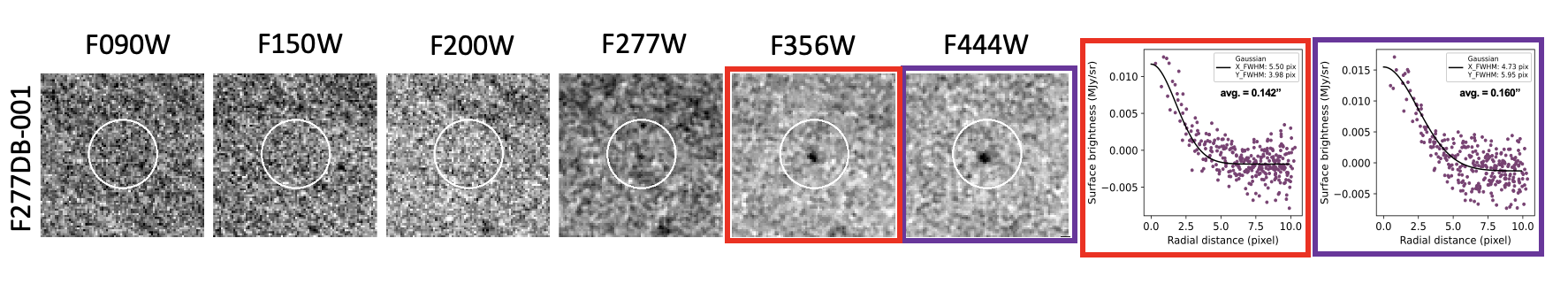}
    \caption{Similar to Figure \ref{fig:pts_f150d}, but for the one point-source
    F277W dropouts in the Y23 sample (among 12 in total). The images used for 
    the profile fitting are outlined by the red (for F356W) and purple (for 
    F444W) boxes.
}
    \label{fig:pts_f277d}
\end{figure*}

\begin{figure*}[htbp]
    \centering
    \includegraphics[width=\textwidth]{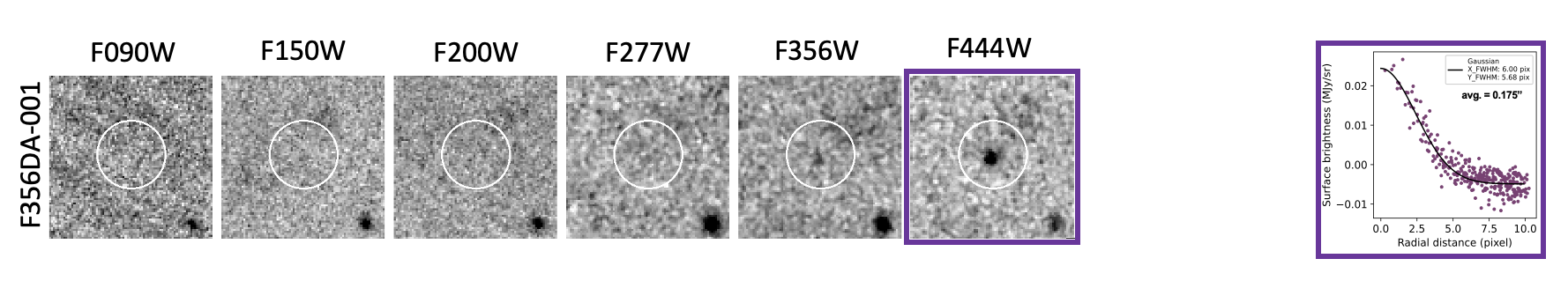}
    \caption{Similar to Figure \ref{fig:pts_f150d}, but for the only object in 
    the new F356W dropout sample truncated at $m_{444}\leq 29.0$~mag. The F444W
    image (outlined by the purple box) of this object satisfies our point-source 
    criteria. 
}
    \label{fig:pts_f356d}
\end{figure*}

   The above criteria identify six point sources among the 60 F150W 
dropouts of Y23. Figure \ref{fig:pts_f150d} shows their 6-band image stamps as
well as the gaussian profile fitting results in the relevant bands. Five of
them have F200W FWHM values sufficient for our point-source criteria, 
among which three also meet the criteria in F277W. The other one satisfies the
criteria in F277W. Among the 15 F200W dropouts of Y23, two are identified as
point sources. Their image stamps and the profile fitting results are shown in
Figure \ref{fig:pts_f200d}. Both of them have F277W and F356W images satisfy
the criteria, one of which also satisfies the criteria in F444W (but this
fitting result is not shown). One of the 12 Y23 F277W dropouts
meets the criteria in both F356W and F444W, which is shown in Figure 
\ref{fig:pts_f277d}. Finally, the only object in the $m_{444}\leq 29.0$~mag
F356W dropout sample satisfies the criteria in F444W, which is shown in
Figure \ref{fig:pts_f356d}. The catalog of all these point-source dropouts is
given in Table \ref{tbl:pts_dropouts}.
We note that the magnitudes of these objects are slightly different from
those quoted in Y23, because we carried out photometry using the new mosaics
based on the updated flux calibration. Following Y23, the matched-aperture
photometry was done by running SExtractor \citep{Bertin1996} in the dual-image
mode to ensure accurate color measurements. For the F356W dropout, the
detection image was the F444W stack. For all other objects, the detection
image was the F356W stack. The background was estimated locally. The isophotal 
magnitudes ``MAG\_ISO'' were adopted; as argued in Y23, the sources of our
interest are small enough in the images such that the MAG\_ISO apertures 
include nearly all the source flux while minimizing the background noise.  

%\begin{longrotatetable}
    \begin{deluxetable*}{cccccccccc}
      \tablecaption{Catalog of point-source dropouts}
      \label{tbl:pts_dropouts}
        \tabletypesize{\footnotesize}
\tablehead{
    \colhead{ID} & \colhead{Short ID} & \colhead{$m_{090}$} & \colhead{$m_{150}$} & \colhead{$m_{200}$} & \colhead{$m_{277}$} & \colhead{$m_{356}$} & \colhead{$m_{444}$}  \\
}
\decimals
\startdata
F150DB J072330.55-732733.12 & F150DB-033 & $>$ 29.01 &   $>$ 29.23    & 28.45$\pm$0.23 & 28.73$\pm$0.10 & 27.95$\pm$0.05 & 27.64$\pm$0.05    \\
F150DB J072324.58-732715.08 & F150DB-050 & $>$ 29.01 &   $>$ 29.23    & 28.44$\pm$0.25 & 28.54$\pm$0.10 & 28.85$\pm$0.12 & 29.20$\pm$0.24    \\
F150DA J072255.88-732917.48 & F150DA-020 & $>$ 29.10 & 29.51$\pm$0.52 & 28.51$\pm$0.18 & 28.57$\pm$0.09 & 28.71$\pm$0.08 & 28.60$\pm$0.11    \\
F150DA J072232.48-732833.23 & F150DA-053 & $>$ 29.10 &   $>$ 29.26    & 28.92$\pm$0.23 & 29.15$\pm$0.12 & 29.44$\pm$0.15 & 29.71$\pm$0.29    \\
F150DA J072239.62-732812.19 & F150DA-066 & $>$ 29.10 & 28.79$\pm$0.31 & 28.02$\pm$0.13 & 28.65$\pm$0.08 & 29.13$\pm$0.11 & 29.26$\pm$0.18    \\
F150DA J072252.78-732741.93 & F150DA-082 & $>$ 29.10 & 29.19$\pm$0.50 & 28.44$\pm$0.21 & 29.39$\pm$0.16 & 29.26$\pm$0.13 & 29.03$\pm$0.16    \\
F200DB J072306.42-732719.88 & F200DB-086 & $>$ 29.01 &   $>$ 29.23    & 28.63$\pm$0.38 & 28.09$\pm$0.09 & 27.72$\pm$0.07 & 27.85$\pm$0.11    \\
F200DA J072243.92-732915.78 & F200DA-033 & $>$ 29.10 &   $>$ 29.26    &   $>$ 29.47    & 26.78$\pm$0.03 & 25.78$\pm$0.01 & 25.46$\pm$0.01          \\
F277DB J072317.55-732825.26 & F277DB-001 & $>$ 29.01 &   $>$ 29.23    &   $>$ 29.43    & 30.80$\pm$0.40 & 29.45$\pm$0.12 & 29.06$\pm$0.12          \\
F356DA J072233.26-732911.14 & F356DA-001 & $>$ 29.10 &   $>$ 29.26    &   $>$ 29.47    &   $>$ 30.57    & 30.26$\pm$0.37 & 27.96$\pm$0.07          \\
\enddata
\tablenotetext{}{The first nine objects are from the dropout sample of Y23,
but with the photometry updated using the new calibrations as in the CRDS context
jwst\_1008.pmap. The last object is an F356W dropout from this current work. The
nomenclature follows Y23's catalog table.
}
\end{deluxetable*}
%\end{longrotatetable}

\section{Supernova interpretation}
\label{sec:SNInter}
    Our current exercise is to fit the SEDs of these point-source dropouts with
spectral sequences of Type Ia supernovae (SNe~Ia) and typical Type IIP 
(SNe~IIP) templates. The SNe~Ia templates are derived from the SALT3-NIR models 
of \citet[][]{Pierel2018, Pierel2022}, while those of SNe IIP are based on Peter 
Nugent's templates \citep[][]{Gilliland1999}. We have applied a simple, 
power-law extrapolations in the wavelength ranges not covered by these
templates.

       For each object, the following $\chi^2$ is minimized:
\begin{equation*}   
\begin{gathered}
\chi^2(m_0, z, t, {\bf p}) \ = \\
 \sum_{\lambda_i}\frac{(m_{\rm obs}(\lambda_i) - m_{\rm SED} (\lambda_i; z, t, {\bf p})+m_{0})^2}{\sigma_{\rm obs}^2(\lambda_i)} + \\ \left(\frac{m_0}{\delta m_0}\right)^2,
\end{gathered}
\end{equation*}
where $m_0$, $z$, and $t$ are the magnitude offset, redshift, and time from the 
optical peak of the templates, respectively, ${\bf p}$ denotes either
the SALT3 $(x_1, c)$ parameters for the SNe~Ia templates or the reddening
$E(B-V)$ for the SNe~IIP templates, $m_{\rm obs}$ and $\sigma_{\rm obs}$ are the
observed magnitudes and errors, and $m_{\rm SED}$ represents the template SED
used for the fits, with $\delta m_0$ characterizing the magnitude dispersion of
the given template. The second term on the right-hand side of the above equation
represents a penalty term for the cases where the observations have significant
deviations from the absolute magnitudes of the SEDs. 

    In such SNe~Ia fits, there are four free parameters for each object:
the time of $B$-band maximum, the magnitude offset from the peak $B$-band 
maximum (a positive offset value means that the template is fainter than the
observed SED), and the SALT parameters $x_1$ and $c$ that control the sub-types 
of SNe~Ia. As we are not fitting cosmological distances,
we simply adopt the $B$-band absolute magnitude of an SN~Ia with $x_1 = 0$ and 
$c=0$ to be $\sim -19.3$~mag, which corresponds to a Hubble Constant of $\sim 
71$~km/s/Mpc. The allowed ranges of $x_1$ and $c$ are $[-3, 3]$ and
$[-0.3, 0.3]$, respectively, which capture all the SN~Ia light curves analyzed 
in \citep{Pierel2022}. We set $\delta m_0$ loosely to 1.0~mag to calculate the 
penalty term that disfavors large magnitude offset from the templates.
For SNe~IIP, we assume the absolute $B$-band 
magnitudes of $M_{\rm B}=-18$~mag at the optical maximum with a scatter of 
$\delta m_0 = 3.0$ mag.
 
\begin{figure*}[htbp]
    \centering
    \includegraphics[width=\textwidth]{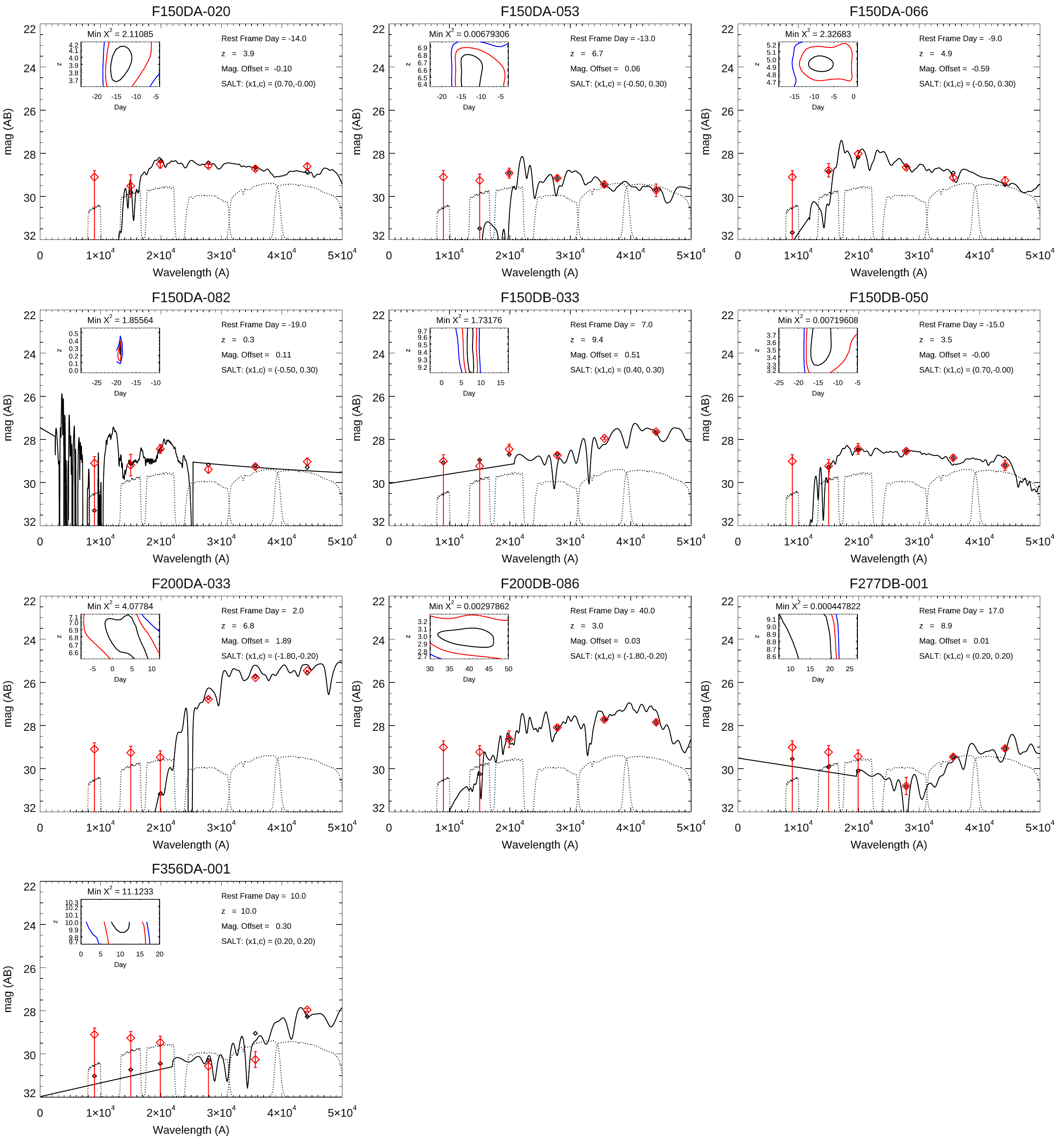}
    \caption{SED fitting results of the point-source dropouts using Type Ia
supernova templates. The object IDs are given at the top of each panel. The red 
symbols show the photometry as in Table~\ref{tbl:pts_dropouts}. The solid lines 
are the best-fit Type~Ia SN templates and the black diamonds are the synthesized 
magnitudes in the six NIRCam bands (indicated by the dotted lines in each panel).
The legends show the time since the optical maximum (a negative value means the 
time before the $B$ maximum) in the rest-frame days, the best-fit redshift, the 
magnitude offset from the best-fit model, and the SALT3NIR $(x_1, c)$ 
parameters. Note that a positive magnitude offset means that observed SED is 
fainter than the template used, as defined by the Equation in 
\S~\ref{sec:SNInter}. The insets show how the redshift and time of SN are 
constrained, with the black, red, and blue contours indicating the 1, 2, and 3 
$\sigma$ levels, respectively. 
}
    \label{fig:SNSED_Ia}
\end{figure*}

\begin{figure*}[htbp]
    \centering
    \includegraphics[width=\textwidth]{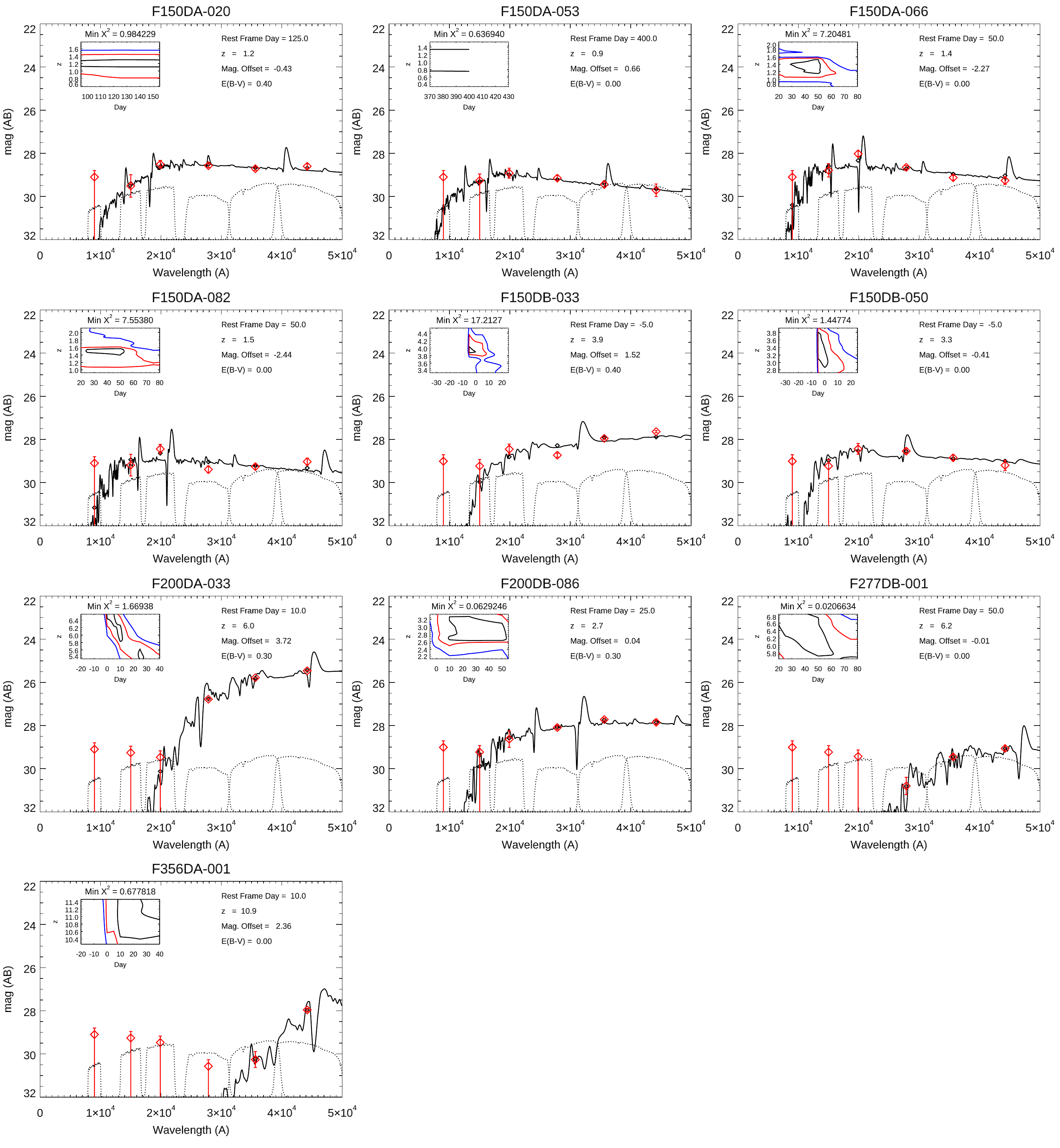}
    \caption{Similar to Figure \ref{fig:SNSED_Ia}, but using Type IIP supernova
templates and with the best-fit $E(B-V)$ value shown in the legend.
}
    \label{fig:SNSED_II}
\end{figure*}

The results are shown in Figure \ref{fig:SNSED_Ia} for SNe~Ia and
Figure \ref{fig:SNSED_II} for SNe~II, respectively. Most candidates agree with 
either type; some of them agree with both but with different best-fit redshifts 
and phases. Specifically, these objects have good fits with SNe~Ia and/or 
SNe~IIP: F150DA-020, F150DA-053, F150DB-050, F200DB-086, and F277DB-001. 
F150DA-033 is consistent with an SN~IIP, but with a large magnitude offset that 
cannot be explained by gravitational magnification by the foreground cluster. 
Likewise, the reddest object, F356DA-001, if it is a transient, cannot be a 
SN~Ia; it can be fitted with the adopted SN~IIP SEDs only if the object is at 
$z \gtrsim 10$. 

    We note two points regarding these fits. First, the magnitude offset is
left as a free parameter in the fitting (even for SNe~Ia), 
and the range of the absolute magnitude distribution we have adopted in our 
fits are broader than normally observed in the local universe. This can be 
justified by the uncertainties introduced by gravitational lensing
magnification and/or interstellar dust extinction.
The objects in module B (indicated by ``B'' in their SIDs) could be amplified by
the gravitational lensing effect due to the cluster
\footnote{However, the amplification
factors quoted in Y23 are not applicable here because these were calculated 
based on the assumption that they are at $z>11$, and therefore these values
cannot be used here directly for correction.}.
The SALT $c$ parameter corrects both the intrinsic SN color and the dust 
extinction. As these are largely based on the observations of local SNe, it is
unclear whether the same corrections are applicable over the redshift range
considered here (for example, the possible interstellar dust extinction at
$z>6$ is highly uncertain). This is exacerbated by the fact that most of our
sources only have positive detections in a few bands, and the wavelength range 
is not broad enough to tightly constrain the interstellar dust extinction. 
In future works with more data (e.g., with new observations at
different epochs), an additional penalty term based on the lensing and reddening 
probabilities can be added to the $\chi^2$ calculation to better account for 
these effects.

    Another point is regarding the $\chi^2$ values. For many of these fits, the
$\chi^2$ values are less than the degrees of freedom (2 and 3 for SNe~Ia and 
SNe~IIP, respectively). This is mainly due to the null detections in the blue 
bands: while they are critical in constraining the redshifts of the objects, 
they are treated as data points with very large errors, which lead to very small 
$\chi^2$. In such cases, only the redshifts are constrained for a given 
template, and the data do not have enough power to constrain the phases and 
intrinsic properties of the SNe. This is shown in the inset contours (for 
example, see the cases for F277DB-001 and F356DA-001). Even with six bands, the 
single-epoch SEDs alone cannot distinguish most SNe~Ia from SNe~IIP (in fact,
from any other types of SNe). For most candidates, both the SNIa and SNIIP can
provide satisfactory fits; the exceptions are F200DA-033 and F356DA-001 which 
strongly favor the SNe~IIP identifications. This is because the redshifts and 
other template parameters are all correlated quantities in the fits. As the 
result, an SN~Ia at a given phase and redshift may be equally well fit by an 
SN~IIP at a different phase and a different redshift. Only at some certain 
phases, the SEDs are different enough that their spectral types can be robustly 
identified based on single-epoch photometry. Had we had multi-epoch 
observations, the SN types and the phases would be much better constrained.

\section{\label{discussion} Discussions}

\subsection{Implications of the supernova interpretation}

   While our result suggests that the point-source dropouts are consistent with
being SNe, by no means it proves that they are. One argument against this 
interpretation is the lack of obvious host galaxies associated with these 
sources. If these are SNe, this implies that they are in very diffuse,
faint galaxies that are not detected by the current NIRCam data. SNe~Ia in 
extremely faint galaxies have been observed by the HST, and it has been 
suggested that SNe~Ia rates may be enhanced in dwarf galaxies or globular 
clusters \citep[][]{Graham2015}. 
Therefore, it is not unreasonable that our objects are hostless SNe. If true,
this may imply that hostless SNe dominate the SN production at z $\sim$ 3.
They can have a profound impact to the metal enrichment of the intergalactic 
medium because their ejecta can escape freely into the intergalactic medium. 
This in fact corroborates with the Chandra X-ray observations of the 
intracluster medium, which shows an early metal enrichment of intracluster 
medium \citep[][]{Mantz2017} as evidenced by a constant metallicity at large 
radii (0.5--1 $r_{500}$) from the cluster center
\footnote{$r_{500}$ is the radius at which the density is 500 times the mean 
density of the universe at the redshift of the cluster} 
and a late-time increase in enrichment at intermediate 
radii (0.1--0.5 $r_{500}$).

   Another concern is whether it is physically plausible to have so many SNe in
such a small field. It is noteworthy that, at the depth of these JWST 
observations, a signficant number of SNe~Ia and even some of the bright SNe~II 
to redshifts as high as $z\approx 6$ can be detected. By extrapolating the local
rate of SNe~Ia following the constraint of the global star formation history of 
the universe, one would expect to discover $2.5\pm 0.6$ SNe~Ia at 
$z\approx 2$--6 in a single NIRCam pointing at the depth of $\sim$30~mag
\citep{WangFLARE2017,RV2019}. 
The number of SNe~II is comparable, although they only extend to a lower 
redshift range as they are generally less luminous. Therefore, one would expect
a total of $\sim$5 SNe in one NIRCam field like what we study here. Given the 
uncertainty of the predictions, this is in broad agreement with the hypothesis 
that at least a large fraction of the point-source dropouts studied here are SNe.

\subsection{Supernovae as possible contaminants in $z>11$ candidate galaxy sample}

    Y23 cautioned that there could be some new types of contaminators in
the $z>11$ candidate search that were not encountered before. This current work 
shows that SNe could be one such type that should be considered if the candidate 
is point-like. Taking the $\chi^2$ at face value, the SNe templates provide
better fits to these objects than in Y23.  Such point-source dropouts are present
in the 1.5--2.0~$\mu$m dropout sample of \citet[][]{Yan2023b} as well, and likely
also in elsewhere. To effectively remove
such contaminants, multiple-epoch imaging is probably the most efficient.
Due to the time dilation at high redshifts ($z\gtrsim 3$ in this context), SNe
will not necessarily manifest themselves as transients in the multiple-epoch
observations. However, they can be singled out as variable objects. 

    Among the ten point-source dropouts considered here, nine are in the Y23 
sample (it does not include any F356W dropouts) that has 87 objects in total. If 
these point sources are indeed SNe, the contamination rate due to SNe is only 
$\sim$10\%. Therefore, the tension between the current NIRCam $z>11$ candidates 
and the previously favored model predictions cannot be removed by resorting to 
this new type of contaminator; the non-point-source candidates cannot be 
explained in this way.

    An interesting point is that a point-source F356W dropout will have to be
at $z>10$ if interpreted as an SN. This is because the rest-frame blue-end 
cutoff at $\sim$4000\AA\ occurring in F356W means that the redshift must be
at $z>10$. This is shown in both Figures \ref{fig:SNSED_Ia} and 
\ref{fig:SNSED_II} for object F350DA-001. As it takes
time for a low-mass star to evolve to a white dwarf that is needed for an SN~Ia,
most likely an SN at such a high redshift cannot be an SN~Ia but should be a 
CCSN. Therefore, ironically, finding point-source F356W dropouts and attributing 
them to SNe would still suggest that the global star formation rate density 
(GSFRD) at $z>10$ must be much higher than what was previously favored. We will 
defer a more detailed calculation to a future paper.

For simplicity, we only consider SNe~Ia and IIP in this work. However, we 
note that other types of SNe could also be used to explain such sources. These 
could include rarer types such as the hydrogen deficient SNe~Ib/c, SNe with 
strong ejecta-circumstellar interaction (SNe~IIn), and super luminous supernovae 
(SLSNe). The theoretically perceived pair-production supernovae (PPSNe) from 
zero-metallicity stars are also a possibility. Acquiring multi-epoch data can
enable further investigations on such scenarios.
   
\section{\label{summary} Summary}

     In this work, we investigate the problem of the point-like sources in the
$z>11$ candidate galaxy sample of Y23, which are unlikely Galactic brown dwarf
stars. We find that such sources might indeed be a new kind of contaminators to 
high-$z$ candidate samples: these could be SNe at various redshifts. This 
somewhat alleviates the tension but does not eliminate it, as there are plenty 
of non-point-source objects in the $z>11$ candidate samples published to date.
As a reference, the point-like sources only constitute $\sim$10\% of the Y23
sample.

    On the other hand, this work shows that SNe at $z>3$ might already have
been detected in the NIRCam data and that they could be singled out using SED 
fitting to point-source dropouts at $>1.5$~$\mu$m.  Finding SNe at $z>3$ will
have a series of important implications. Multiple-epoch 
NIRCam imaging is the most efficient way to test the SN hypothesis. Due to the 
time dilation and the high sensitivity of NIRCam, SNe at high redshifts most 
likely would show up as variable objects (but not necessarily as transients) in 
multiple-epoch NIRCam images. Such data would greatly improve the constraints on 
their redshifts and the elapsed time before/after the maximum. This will enable
quick identifications of the most important SN candidates, e.g., SNe~Ia at 
$z\approx 6$--10, for spectroscopic confirmation.

 Finally, we remark that SNe~Ia at such redshifts can be used to quantitatively
constrain the systematic redshift-evolution of the intrinsic properties of
SNe~Ia \citep{Riess_2006,LuJia} by incorporating explicit redshift-dependent
light curve shape and color corrections to SN~Ia magnitudes.

All the {\it JWST} data used in this paper can be found in MAST: 
\dataset[10.17909/7rjp-th98]{http://dx.doi.org/10.17909/7rjp-th98}.

\begin{acknowledgements}
HY acknowledges the partial support from the University of Missouri 
Research Council Grant URC-23-029. LW acknowledges support from the NSF through 
the project AST-1817099. LH acknowledges support from China Postdoctoral Science 
Foundation (grant no. 2022M723372) and the Jiangsu Funding Program for Excellent 
Postdoctoral Talent.
\end{acknowledgements}

\appendix
\setcounter{figure}{0}
\renewcommand{\thefigure}{A\arabic{figure}}

\begin{figure*}[htbp]
    \centering
    \includegraphics[width=\textwidth]{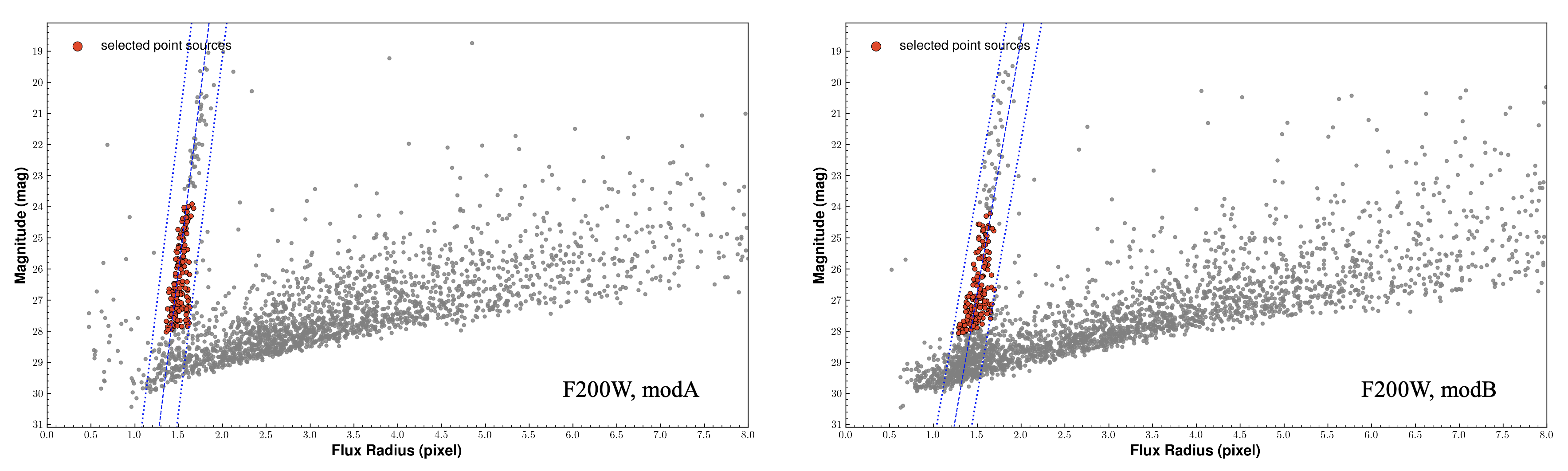}
    \caption{Selection of bright point sources to measure the FWHM distribution,
    using F200W as an example. The  left panel is for module A and the right panel
    is for module B, respectively. The Y axis is the SExtractor MAG\_AUTO magnitude,
    while the X axis is the  radius (in pixels; 30mas pixel scale) of a circle 
    aperture that contains half of the total flux.
    The point sources are within a narrow ``belt'' as outlined. The red dots are
    the sources retained for the statistics. 
    }
    \label{fig:ptselection200w}
\end{figure*}

\begin{figure*}[htbp]
    \centering
    \includegraphics[width=\textwidth]{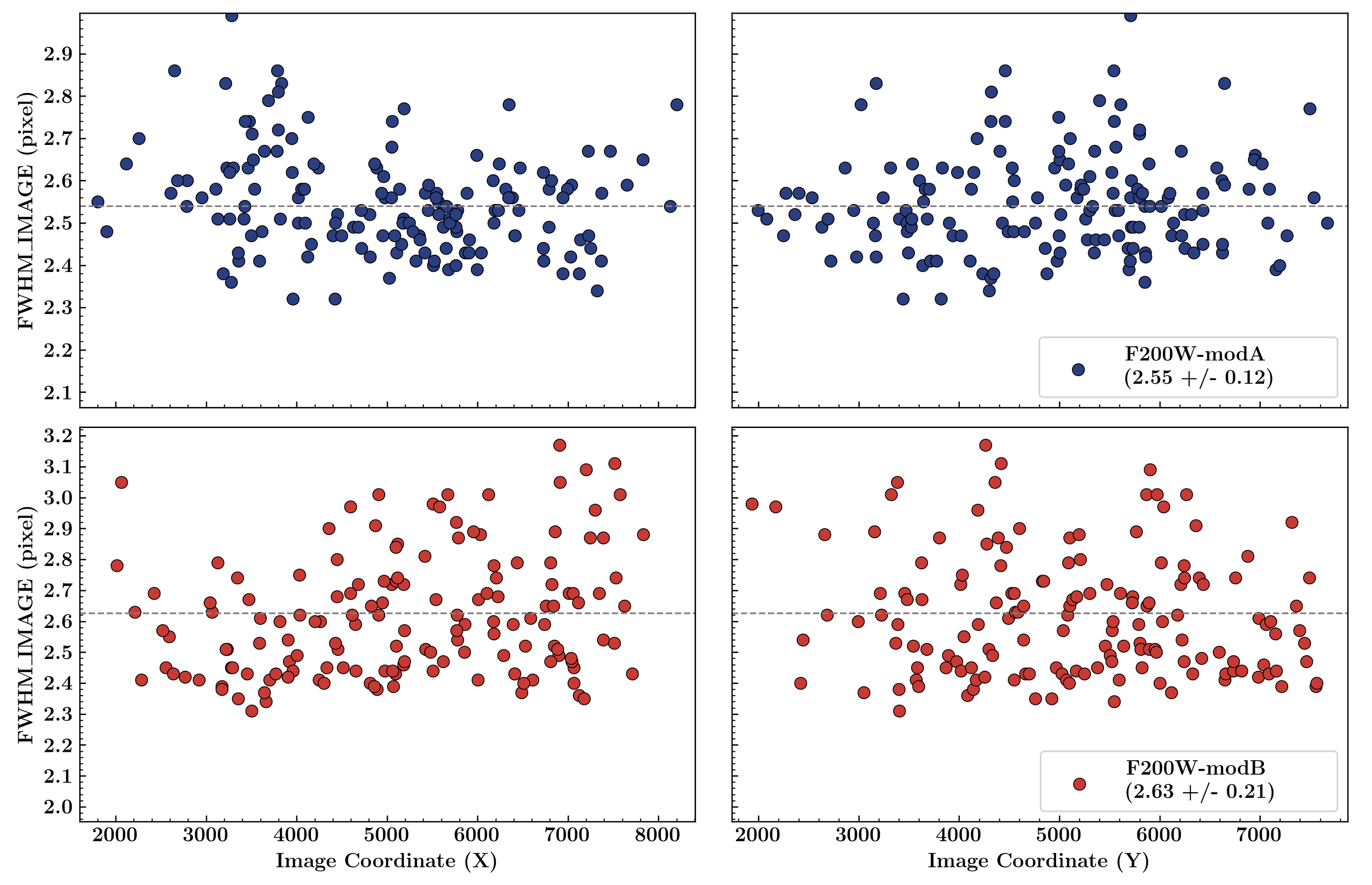}
    \caption{FWHM distribution of bright point sources as a function of image
    coordinates (X in left and Y in right, respectively). The case in F200W is shown
    here as an example. The top panels are for module A and the bottom panels are for 
    module B, respectively. The FWHM values are in pixels; the pixel scale is 30mas.
    }
    \label{fig:ptfwhm200w}
\end{figure*}

To determine the point source FWHM distributions, we used the methodology of
the PSFEx tool \citep{Bertin2011}. In the magnitude (SExtractor's MAG\_AUTO) 
versus half-light radius (SEXtractor's FLUX\_RADIUS with PHOT\_FLUXFRAC~$=0.5$)
plot, point sources occupy a nearly vertical locus. This is shown in Figure
\ref{fig:ptselection200w} for our case in F200W as an example.
The initial point sources were selected within this locus, with the additional 
constraints that they should have high S/N (SExtractor's SNR\_WIN~$> 30.0$) and 
be relatively round (ELLIPTICITY~$< 0.3$). We ran PSFEx on these initial sources
and measured the average FWHM after 5~$\sigma$ 
clipping. The module A (B) FWHM values and rms thus obtained are 
0\arcsec.076 $\pm$ 0\arcsec.004, 0\arcsec.121 $\pm$ 0\arcsec.005, 
0\arcsec.139 $\pm$ 0\arcsec.004, 0\arcsec.161 $\pm$ 0\arcsec.006
(0\arcsec.079 $\pm$ 0\arcsec.006, 0\arcsec.124 $\pm$ 0\arcsec.007,
 0\arcsec.142 $\pm$ 0\arcsec.005, 0\arcsec.164 $\pm$ 0\arcsec.004)
in F200W, F277W, F356W, and F444W, respectively.
The FWHM distribution of the retained point sources in F200W is shown in Figure
\ref{fig:ptfwhm200w} as an example.

\newpage

%\bibliography{myref}
\bibliographystyle{aasjournal.bst}

\end{document}